\documentclass[aps,pre,twocolumn,groupedaddress]{revtex4-1}

\usepackage{graphicx}
\usepackage{dcolumn}

\usepackage{bm}
\usepackage{amsfonts}
\usepackage{amsmath}

\usepackage[dvips]{color}

\def\mrm{\mathrm}
\def\mbf{\mathbf}

\def\qq{\mbf{q}}
\def\rr{\mbf{r}}

\def\AA{\mbf{A}}
\def\qGS{\mbf{q}_\mrm{GS}}

\def\mrm{\mathrm}
\def\mbf{\mathbf}

\def\qq{\mbf{q}}
\def\rr{\mbf{r}}

\def\round{\partial}
\def\rot{\mrm{rot}}

\def\vphi{\varphi}
\def\phit{\tilde{\phi}}

\def\pphi{\bfs{\phi}}
\def\pphit{\tilde{\bfs{\phi}}}
\def\vvphi{\bfs{\varphi}}

\newcommand{\bfs}[1]{\boldsymbol{#1}}

\begin{document}


\title{
Transition by Breaking of Analyticity in the Ground State of Josephson Junction Arrays 
as a Static Signature of the Vortex Jamming Transition
}



\author{Tomoaki Nogawa}
\email{nogawa@serow.t.u-tokyo.ac.jp}
\affiliation{%
Department of Applied Physics, 
The University of Tokyo, Hongo, Bunkyo-ku, Tokyo 113-8656, Japan
}
\author{Hajime Yoshino}
\affiliation{%
Department of Earth and Space Science, 
Faculty of Science, Osaka University,
Toyonaka 560-0043, Japan
}%
\author{Bongsoo Kim}
\affiliation{%
Department of Physics, Changwon National University, 
Changwon 641-773, Korea
}%



\date{\today}

\begin{abstract}
We investigate the ground state of the irrationally frustrated Josephson junction array 
with controlling anisotropy parameter $\lambda$ 
that is the ratio of the longitudinal Josephson coupling to the transverse one. 
We find that the ground state has one dimensional periodicity 
whose reciprocal lattice vector depends on $\lambda$ 
and is incommensurate with the substrate lattice. 
Approaching the isotropic point, $\lambda$=1 
the so called hull function of the ground state exhibits analyticity breaking 
similar to the Aubry transition in the Frenkel-Kontorova model. 
We find a scaling law for the harmonic spectrum of the hull functions, 
which suggests the existence of a characteristic length scale diverging at the isotropic point.
This critical behavior is directly connected to the jamming transition 
previously observed in the current-voltage characteristics by a numerical simulation.
On top of the ground state there is a gapless, continuous band of metastable states, 
which exhibit the same critical behavior as the ground state.
\end{abstract}


\maketitle

\section{Introduction}

Frustration is regarded as a key concept in the understanding of a number of complex cooperative phenomena in condensed matter systems including the glass transition \cite{Tarjus05, Sadoc99}. 
In general frustration leads to flattening of the energy landscape 
and suppresses the onset of conventional long-range orders, 
thereby opening possibilities of exotic phases. 
In order to understand the roles of frustration, 
it is desirable to vary the strength of frustration in a systematic way.
The frustrated Josephson junction array (JJA) under an external magnetic field \cite{Tinkham04} 
provides an ideally simple setting for this purpose 
since the strength of the frustration in the JJA can be tuned at will 
as discussed below.
The origin of frustration of this system is the impossibility 
to minimize the Josephson junction coupling energy of longitudinal and transverse bonds 
at the same time when a magnetic field is applied.
This mechanism seems to have a certain generality.
The frustrated JJA is closely related to the notion of frustrated
crystals in curved space \cite{Tarjus05, Sadoc99}.
Such frustrated crystals may be realized, to some extent, by actually 
bending crystals \cite{Nogawa06, Hayashi06}. 
This amounts to injecting a given number density of
dislocations into the crystals.
Here the key parameter of the frustration 
is the number density of the externally induced dislocations.
In the frustrated JJA, the dislocations correspond to the vortices
induced by an external magnetic field. 
Quite remarkably, one can easily control the number density of vortices per plaquette $f$ 
by just changing the strength of the applied external magnetic field.

By choosing an irrational $f$ instead of a rational $f$ \cite{Straley93} 
we can maximize the strength of frustration
in the sense that we can suppress the formation of a dislocation (vortex) lattice
commensurate with respect to the underlying JJA.
This system is called an {\it irrationally frustrated} JJA (IFJJA) 
and has attracted researchers for a long time.
It has been observed numerically that the system remains 
in a vortex liquid state down to low temperatures 
\cite{Park00, Granato07, Granato08} 
and exhibits glassy signatures \cite{Halsey85l, Kim97}.

Recently it has been found that a relevant 
and very interesting perturbation on the IFJJA 
is the {\it anisotropy} of the Josephson coupling 
\cite{Yoshino09,Yoshino10,Yoshino10p}, 
which breaks the balance of longitudinal and transverse bonds 
and thus weakens the frustration. 
Anisotropic JJAs can be fabricated in laboratories by the lithography technique \cite{Saito00}. 
The IFJJA on the square lattice with different strengths of the Josephson couplings 
$J_{x}$ and $J_{y}$ in the $x$ and $y$ directions, manifests itself as unusual vortex matter 
that slides freely in the direction of stronger coupling even at zero temperature, similarly to incommensurate charge density waves,  but jammed into the direction of weaker coupling \cite{Yoshino09}.
It was argued  that the mechanism of the sliding-jamming transition is similar to that of the Frenkel-Kontorova (FK) model \cite{Peyrard83}.

In the present paper, we study how the ground states (GSs) of the IFJJA change with the anisotropy $\lambda=J_{y}/J_{x}$.
To this end we develop an efficient numerical method valid even up to the isotropic point $\lambda=1$, which substantially extends the previous analysis based on a $1/\lambda$ expansion \cite{Yoshino10}, whose validity is limited to the strong anisotropy limit $\lambda \gg 1$. 
The GSs are characterized by an incommensurate wave vector ${\bf q}(\lambda)$ 
and its higher harmonics that continuously vary with $\lambda$. With strong enough anisotropy $\lambda \gg 1$, the vortices are aligned as stripes parallel to the weaker coupling axis
\cite{Yoshino10}. 
By tuning the anisotropy smaller, the stripes become more tilted 
with respect to the weaker coupling (see Fig.~\ref{fig:vortex}).
We demonstrate that the so called hull function \cite{Peyrard83}, 
which reflects the hidden periodic pattern that is incommensurate 
with the substrate lattice, becomes {\em nonanalytic} 
at the isotropic point $\lambda=1$. 
Physically this means that
the sliding becomes prohibited in both the $x$ and $y$ directions 
at the isotropic point, leading to a jamming transition, 
which is analogous to the so-called Aubry transition, 
which is well known in the FK model \cite{Peyrard83}.
This analyticity-breaking transition is thus a static manifestation 
of the  observed jamming transition.


 To study the IFJJA with anisotropic Josephson couplings,
we consider a classical model described by the following Hamiltonian with the 
Josephson coupling between nearest neighbors along the $x$ and $y$ axes: 
\begin{equation}
H(\lambda, \{ \pphi_i \} )= -\sum_i \Big[ \cos \phi_i^x + \lambda \cos \phi_i^y \Big]
\label{eq-jja}
\end{equation}
with the gauge-invariant phase differences across the junctions, 
\begin{equation}
\pphi_i = (\phi_i^x, \phi_i^y) \equiv (
\theta_{i+\hat{x}}-\theta_{i}-A_i^x, 
\theta_{i+\hat{y}}-\theta_{i}-A_i^y
)
\label{eq:phi-theta}
\end{equation}
Here $\theta_{i}$ ($i=1,2,\ldots,N$) is the phase of superconducting order parameter 
on the $i$-th island located at the lattice point $\rr_{i}=(x_{i},y_{i})$ 
of a square lattice of size 
$N=L\times L$ with $0 \le x_i <L$, $0 \le y_i < L$,  
$\rr_{i+\hat{x}}=(x_{i}+1,y_{i})$ and $\rr_{i+\hat{y}}=(x_{i},y_{i}+1)$.
The parameter $\lambda$ denotes the strength 
of the coupling anisotropy. 
Due to the symmetry for permutation of axes $x$ and $y$, 
it is sufficient to investigate only $\lambda \ge 1$. 
The presence of the external magnetic field is described by the vector potential
$\AA_i = (A^x_i, A^y_i)$.
It is related to the filling factor $f$, 
which is the average number of flux quanta per square plaquette, 
by taking its {\it lattice rotation} 
$\rot \AA_i = 2\pi f$ with 
$\rot \mbf{X}_i \equiv X_i^x + X_{i+\hat{x}}^y - X_{i+\hat{y}}^x - X_i^y$.

Specifically we employ a commonly used irrational number 
$(3-\sqrt{5})/2 \approx 0.382$ for $f$ \cite{Halsey85l}. 
We believe, however, that the properties we discuss below do not depend on the
specific choice of the irrational number.
To impose a periodic boundary condition, 
$f$ must be approximated by a rational number.
It is known that a good approximation is obtained 
by using the Fibonacci series 
$\{ F_n : F_{n+2}=F_{n+1}+F_n\}$ 
as $f=F_{n-2}/F_{n}$. 
We use two series $F_n=\cdots,8,13,21,34,\cdots$ 
and $F'_n=\cdots, 18,29,47,76,\cdots$ 
corresponding to different initial conditions 
$(F_0=1, F_1=2)$ and $(F'_0=1, F'_1=3)$. 
We use system sizes $L=F_n$ to ensure truly irrational filling 
in the thermodynamic limit $L \to \infty$.

\section{Variation analysis}

\begin{figure}[t]
\begin{center}
\includegraphics[trim=12 10 12 10,scale=0.330,clip]{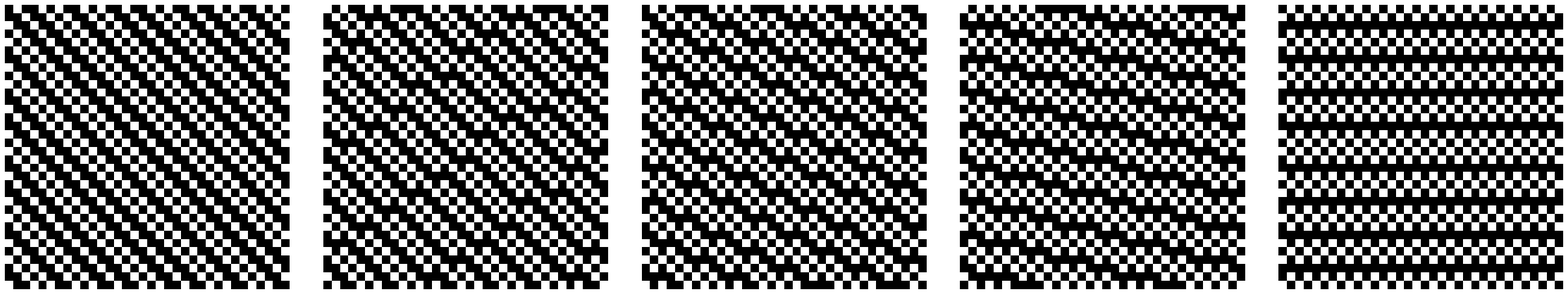}
\end{center}
\vspace{-5mm}
\caption{\label{fig:vortex}
Vortex configurations of the metastable states (candidates for the GSs) 
with various FRLV $\qq=(q^x,f)$ in the anisotropic IFJJA with $\lambda \geq 1$.
White squares indicate the plaquettes containing vortices. 
The panels from left to right correspond, respectively, to 
$q^x L=$13,14,15,16, and 17 with $L=34$.
As the anisotropy $\lambda$ increases, 
the GS changes from the left to right.
Note that there is no recursive unit except the $L \times 1$ strip 
and no simple relation between the tilt of apparent vortex stripes and the FRLV.
}
\end{figure}

We now study the GS of the system by extending the previous analysis 
that was limited to the computation of a few terms in the $1/\lambda$ expansion  \cite{Yoshino10}.
First we performed numerical searches of the GSs at various strengths of 
the anisotropy $\lambda$ on small systems using a simulated annealing method 
(not shown here). 
We found that the vortex configurations in the GSs exhibit stripe patterns 
whose tilt angle with respect to the axis of the weaker coupling 
varies with $\lambda$ (see Fig.~\ref{fig:vortex}). 
This comes from the fact that $\pphi_i$ has {\it one-dimensional} periodicity, 
which is generally incommensurate with the underlying JJA, 
and a fundamental reciprocal lattice vector (FRLV) $\qq(\lambda)$ 
varies with $\lambda$. 
We utilize this fact in the following analysis. 
In the large anisotropy limit it was found to be 
$\qGS(\infty)=(1/2, f)$ (shown in the right most panel 
in Fig.~\ref{fig:vortex}) \cite{Yoshino10}.

Let us now explain our strategy, which is slightly reformulated 
but equivalent to the scheme presented in Ref.~\cite{Yoshino10} 
except for some partial use of numerical procedures.
First, recalling Eq.~(\ref{eq:phi-theta}), 
we find that the phase differences must satisfy the following equation at each plaquette: 
\begin{equation}
\mrm{rot} \pphi_i = -\rot \AA_i = -2 \pi f. 
\label{eq:monovalue}
\end{equation}
To meet this condition, it is convenient to decompose $\pphi$ into 
a uniform-rotation part and a rotation-free part as 
\begin{eqnarray}
\pphi_i = \pphi^*_i + \sum_{n = 1}^{n_\mrm{max}-1} \vvphi_{n} 
e^{2 n \pi i \qq \cdot \rr_{i} }. 
\end{eqnarray}
where 
\begin{eqnarray}
\rot \pphi^*_i = -2\pi f
\quad
\rot \left[ \vvphi_n e^{2 n \pi i \qq \cdot \rr_{i} } \right] = 0. 
\label{eq:rot}
\end{eqnarray}
The upper bound of the harmonics $n_\mrm{max}$ is 
given by the smallest $n$ with which both $nq^x$ and $nq^y$ are integers. 
It almost always equals $L$ in the present model 
and becomes infinite in the irrational limit. 
The first of Eqs.~(\ref{eq:rot}) can be solved by 
$\pphi^*_i = (-2\pi f y_{i} + C_\qq,0)$, 
where $C_\qq$ is a constant introduced to satisfy the periodic boundary condition 
for $\theta_{i}$.
The second of Eqs.~(\ref{eq:rot}) can be solved by 
$\vvphi_n = (\vphi_n^x, \vphi_n^y)$ with 
$\vphi^y_n/\vphi^x_n = (1-e^{2 n \pi i q^y})/(1-e^{2 n \pi i q^x})$, 
which can be used to eliminate the degrees of freedom 
$\{ \vphi^y_n \}$ in favor of $\{ \vphi^x_n \}$ (or vice versa).

We assume that the FRLV can be parametrized as $\qq=(q^x, f)$ by a parameter $q^x$,
 which is suggested by the $1/\lambda$ expansion analysis \cite{Yoshino10}. 
This perturbative analysis suggests that a unique stable solution, 
which satisfies the current conservation (force balance) condition,
can be constructed explicitly for a given $q^x$ that is left as a control parameter.
The analytic computation is, however, quite cumbersome and difficult to continue to
higher orders. To overcome the difficulty, 
we numerically search the configuration with minimal energy, 
\begin{eqnarray}
E_m(\lambda,\qq) = \min\limits_{\{ \vvphi_n \} } 
H(\lambda, \qq, \{\vvphi_n\}),
\end{eqnarray}
for a given FRLV. We performed this analysis in the range $f \le q^x \le 1/2$.

For numerical energy minimization, we employ the steepest-descent method 
for the degrees of freedom $\{ \vphi^x_n \}$. 
Actually, we numerically integrate the equations of motion;
$d\vphi^x_n/dt=-\round H/\round \vphi^x_n$ for $1 \le n < n_\mrm{max}$, 
with an initial condition $\vphi^x_n=0$ for all $n$. 
We determine that the system relaxes to the minimum-energy state 
when $|dH/dt|$ becomes less than $10^{-6}$.
As an example we show in Fig.~\ref{fig:vortex} 
the vortex configurations corresponding to the 
candidate FRLVs for $L=34$. 

The final step is to determine $\qGS(\lambda)$: $\qq$ 
which gives the minimal energy for given anisotropy $\lambda$ as 
\begin{eqnarray}
E_\mrm{GS}(\lambda) 
= \min\limits_\qq E_m(\lambda, \qq)
= E_m \left(\lambda, \qGS(\lambda) \right). 
\end{eqnarray}

Although there is no rigorous proof that the states obtained 
by this method are the true ground states, 
we find that they coincide with the ones obtained 
by the simulated annealing methods. 
We thus believe that they are the true ground states.

\section{Results}

\subsection{successive structure transition}

\begin{figure}[t]
\begin{center}
\includegraphics[trim=0 20 120 0,scale=0.340,clip]{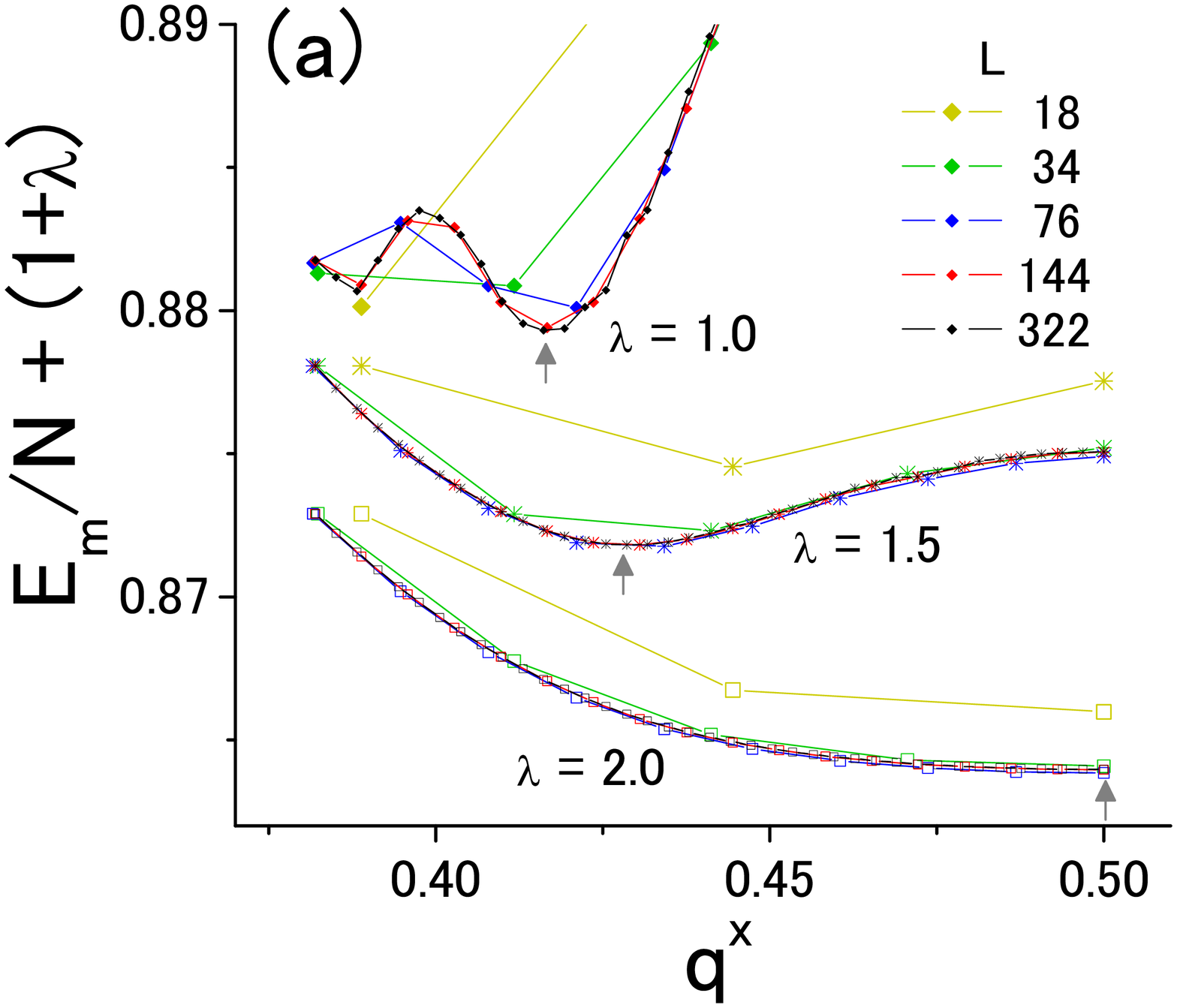}
\\
\includegraphics[trim=0 20 120 0,scale=0.340,clip]{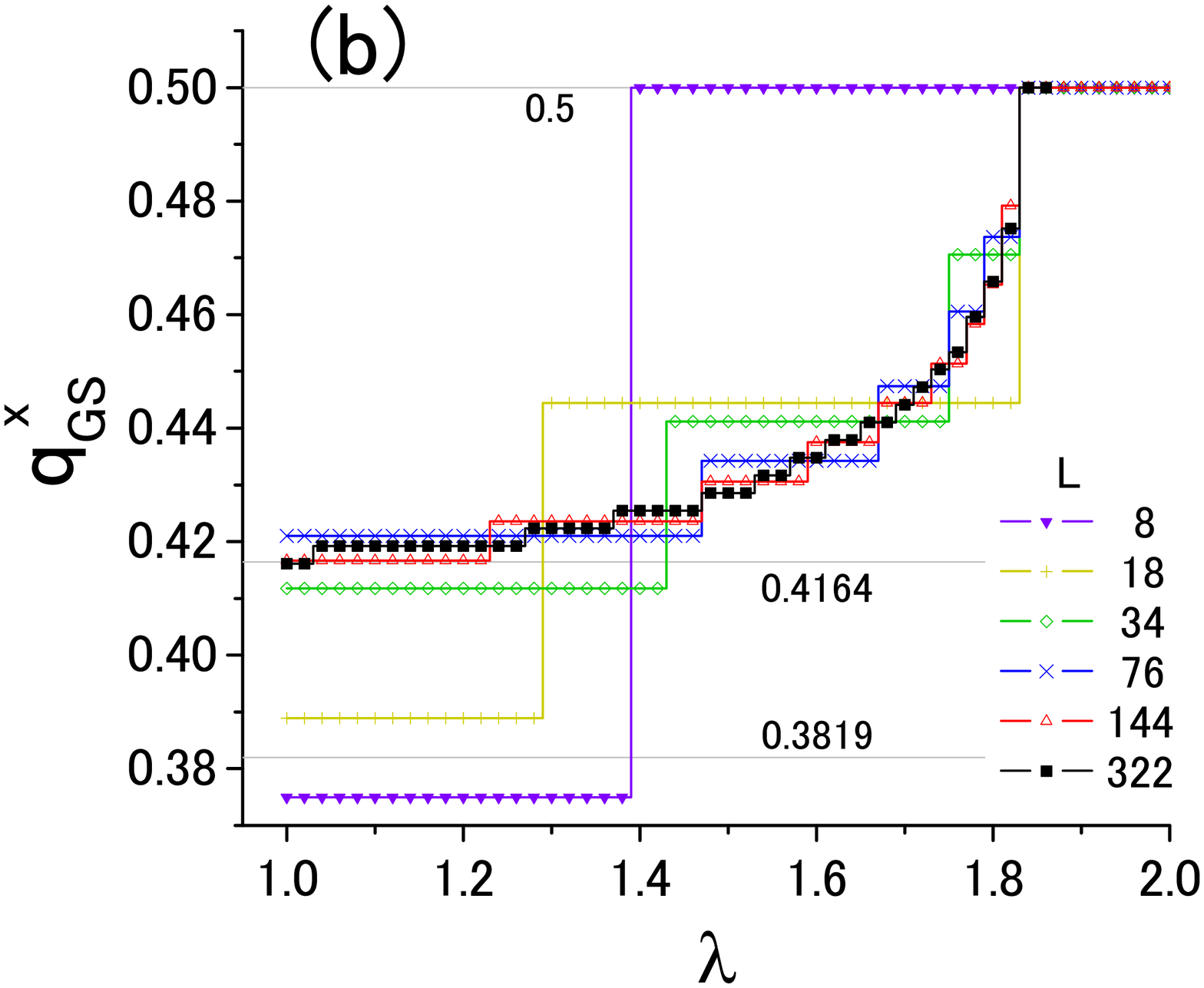}
\end{center}
\vspace{-5mm}
\caption{\label{fig:dispersion}
(Color online) 
(a) Minimal energy vs $q^x$ for various values of $\lambda$.
Constant values are added in the cases of 
$\lambda=1.0$ and $1.5$ (by 0.155 and 0.050, respectively) for 
a convenient view.
(b) Anisotropy dependence of the FRLV of the GS, $\qGS$. 
}
\end{figure}

Figure~\ref{fig:dispersion}(a) shows the minimal energy 
$E_m(\lambda,\qq)$ as a function of $q^{x}$ 
for $\lambda$=1.0, 1.5, and 2.0.
From this we determine $\qGS$ and $E_\mrm{GS}$ 
for a given anisotropy $\lambda$. 
The energy behaves quadratically in the vicinity of each minimum as 
$E_m(\lambda,\qq) - E_\mrm{GS}(\lambda) 
= L^2 c(\lambda) [ \qq - \qGS(\lambda) ]^2$, 
where $c(\lambda)$ is a constant of $O(L^0)$.
Thus there is a continuous band of metastable states 
with slightly different FRLVs around the GS.
(We confirmed that the obtained metastable states are stable 
even if we lift the constraint; the Fourier components equal zero 
except for the harmonics $n\qq$.)

Figure~\ref{fig:dispersion}(b) shows the $\lambda$ dependence of the $x$ component
of the $\qGS(\lambda)=(q_\mrm{GS}^{x},f)$ with $\alpha=x,y$. 
It can be seen that $q_\mrm{GS}^x$ monotonically increases with $\lambda$ 
and changes in a stepwise manner at several points, 
where level crossings occur between metastable states 
with neighboring FRLVs. 
As $L$ becomes larger, the number of steps increases 
and $q_\mrm{GS}^x$ tends to be a continuous function of $\lambda$. 
It can be seen that the horizontal stripe state $\qq=(1/2,f)$ \cite{Yoshino10} 
is no longer the GS for $\lambda \le 1.8$. 
The vortex stripes in the GS become more tilted with respect to the weaker coupling axis 
when approaching the isotropic point $\lambda=1$ (see Fig.~\ref{fig:vortex}).


The GS of the isotropic system ($\lambda=1$) is of particular interest. 
For small sizes $L \leq 18$, the GS is 
the previously found staircase state \cite{Halsey85b, Gupta98, Lee01} 
with $q^x_\mrm{GS}=q^y_\mrm{GS}=f$.
However, this is not the case for $L \geq 34$. 
A new minimum appears at $q^x \approx 0.416$ 
and there is a local minimum at $q^x \approx 0.388$,
both of which are larger than $f=0.381966\cdots$. 
We obtain $q^x_\mrm{GS}$=14/34, 23/55, 37/89, 60/144, and 97/233, 
where the numerators constitute a Fibonacci series 
(with $F''_0=1$ and $F''_1=4$) 
and the denominators correspond to $L$. 
Therefore, $q^x_\mrm{GS}$ seems to equal $0.4164078\cdots$ 
$[ > (3-\sqrt{5})/2 ]$ in the limit of $L \rightarrow \infty$.
A notable feature is that the symmetry for the permutation 
of $x$ and $y$ axes spontaneously breaks 
in contrast to the symmetric staircase state mentioned above.

\subsection{breaking of analyticity in hull functions}

\begin{figure}[t]
\begin{center}
\includegraphics[trim=0 40 120 20,scale=0.340,clip]{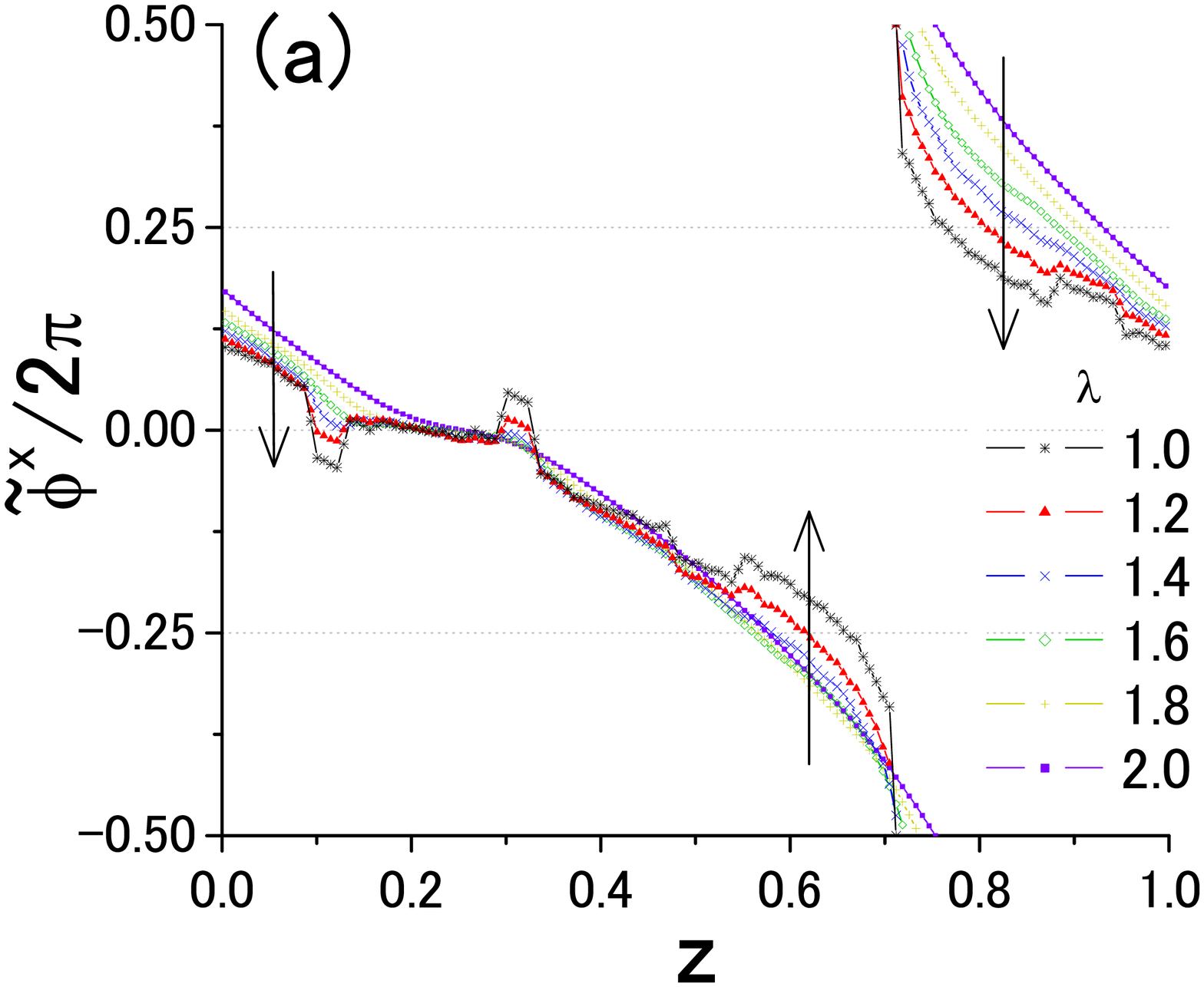}
\\
\includegraphics[trim=0 40 120 20,scale=0.340,clip]{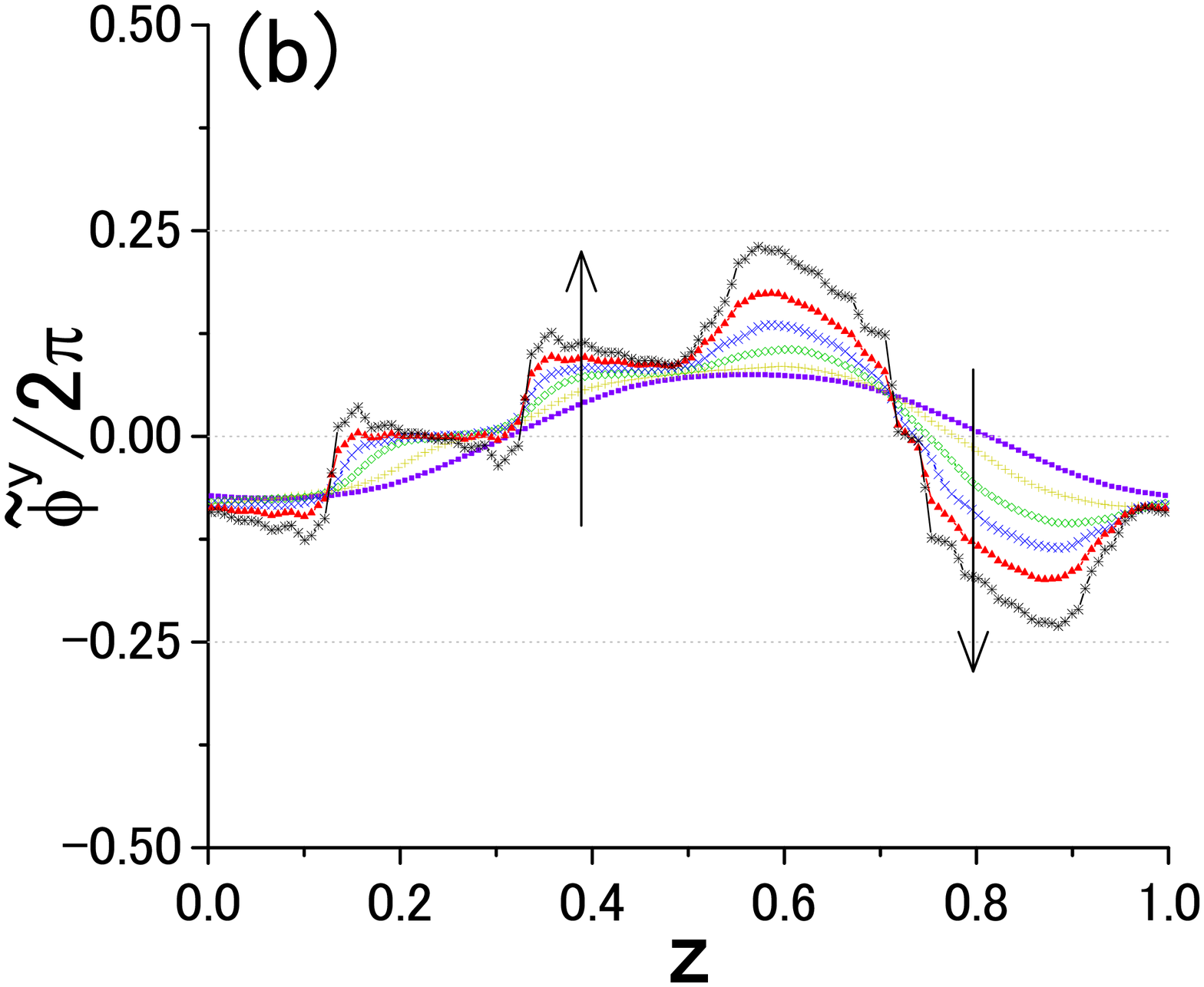}
\end{center}
\vspace{-5mm}
\caption{\label{fig:hull}
(Color online) 
Hull function of (a) $\phi^x$ and (b) $\phi^y$ 
for six different values of $\lambda$.
Each graph has $L$(=144) points. 
The arrows indicate the direction of decreasing $\lambda$. 
}
\end{figure}

Next we investigate more closely the properties of the GSs parametrized by the FRLVs $\qq$, 
which are generally incommensurate with respect to the underlying JJA.
A useful measure of such an incommensurate object is the so called 
hull function \cite{Peyrard83, Floria96, Yoshino10p} 
$\tilde{\pphi}(z)=(\phit^{x}(z),\phit^{y}(z))$, 
with which the phase field is written as 
\begin{equation}
\pphi_i = \pphit(\qq \cdot \rr_i).
\end{equation} 
In Fig.~\ref{fig:hull} the hull functions $\phit^x(z)$ and $\phit^y(z)$ 
are plotted with respect to $0<z<1$. 
It is noteworthy that $\phit^x$ is distributed, 
covering all phase from $-\pi$ to $\pi$, 
while $\phit^y$ is bounded around zero to have a gap around $\phit^x=\pm \pi$. 
When the hull function is smooth and gapless, 
there is a sliding soft mode \cite{Yoshino10}; 
the infinitesimal increment of $\phi^x$ is compensated for 
by the shift in the argument of the hull function $\qq \cdot \rr_i$, 
which costs no energy barrier.
Since this leads to dissipative vortex flow driven by current induction, 
the GSs have superconductivity only along the stronger-coupling
direction. 

\begin{figure}[t]
\begin{center}
\includegraphics[trim=50 70 30 0,scale=0.330,clip]{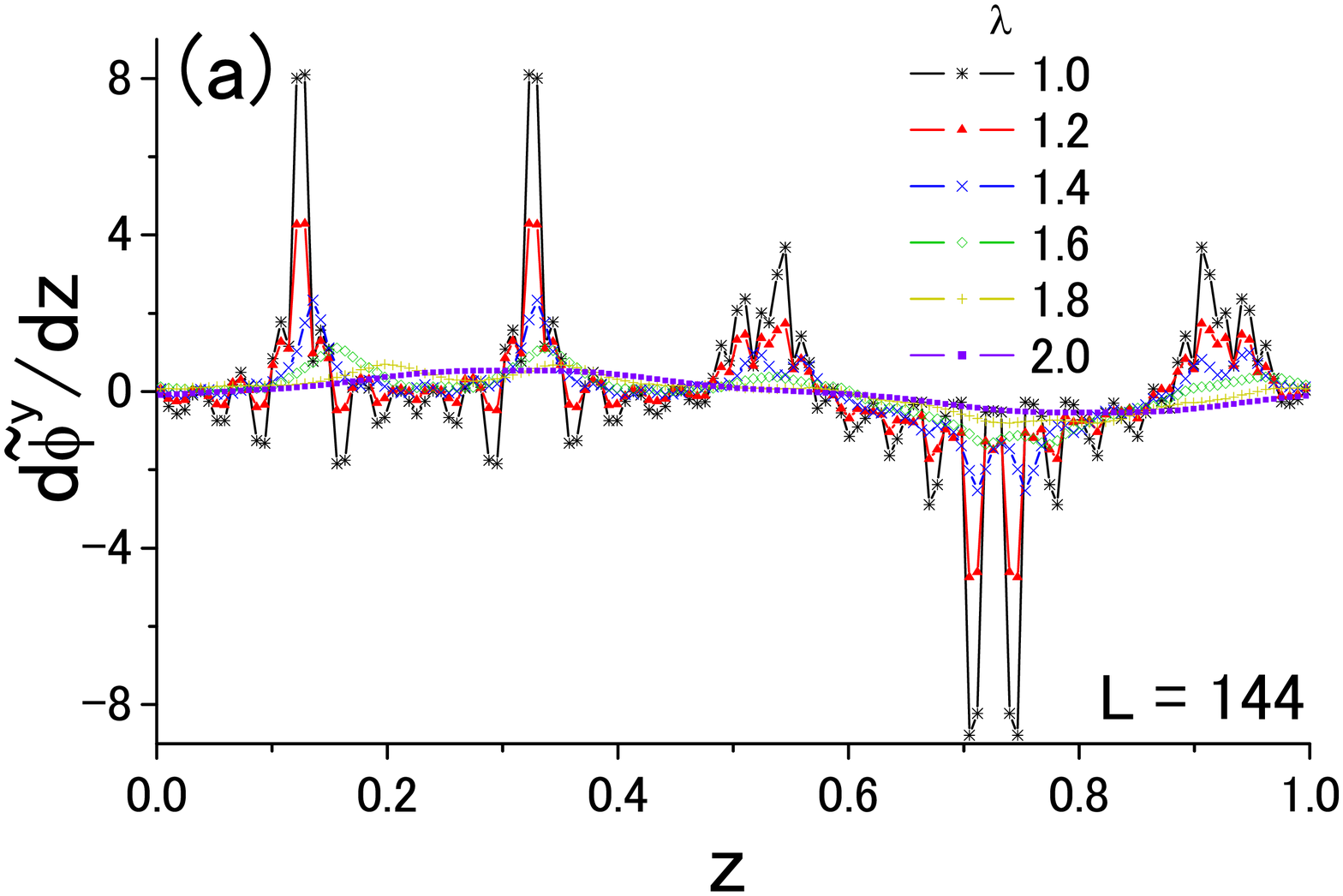}
\\
\includegraphics[trim=50 70 30 0,scale=0.330,clip]{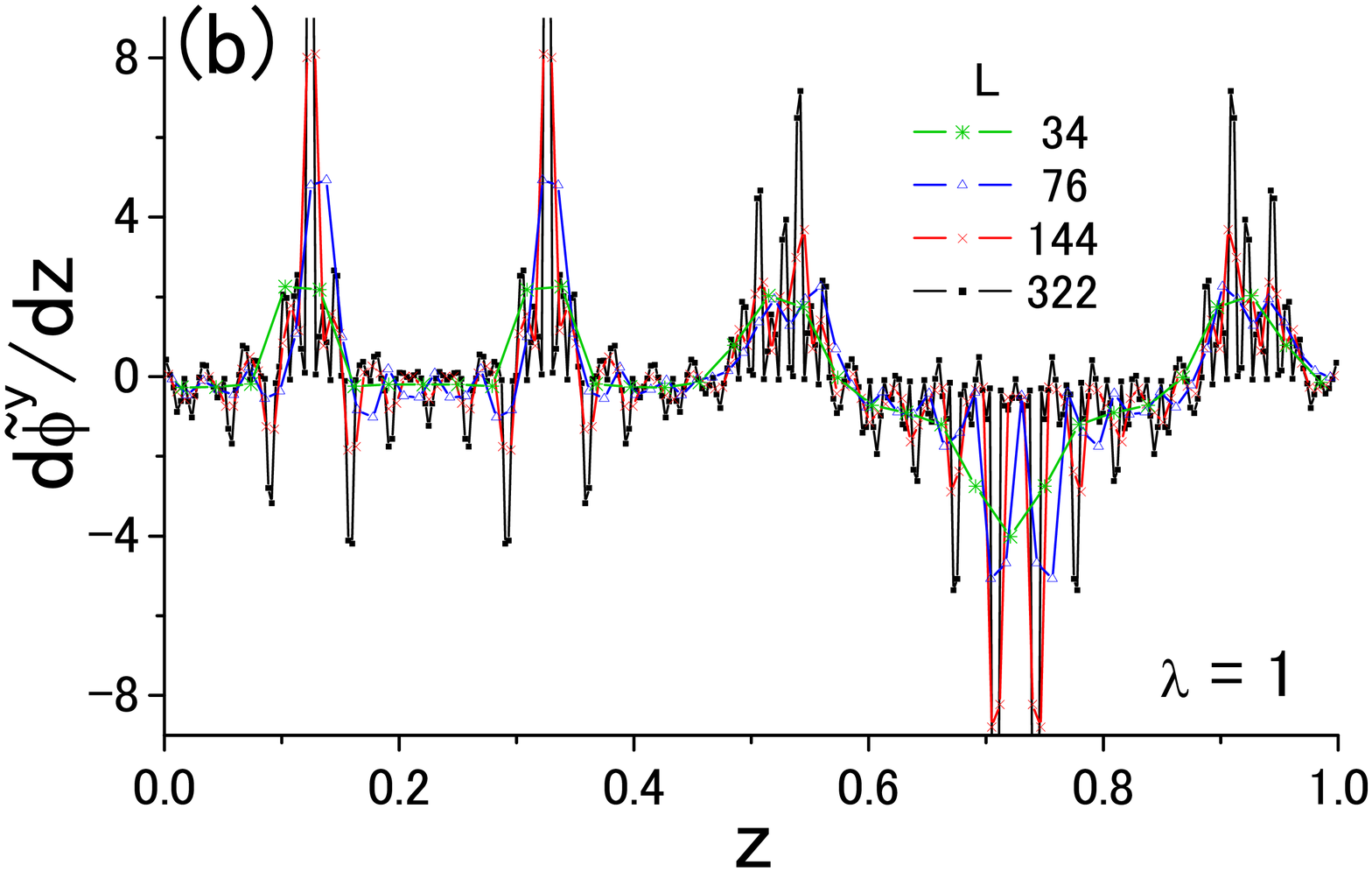}
\end{center}
\vspace{-5mm}
\caption{\label{fig:dphi}
(Color online) 
Derivative of the hull function of $\phi^y$.
(a) Anisotropy $\lambda$ dependence for $L=144$. 
(b) Size $L$ dependence at the symmetric point $\lambda=1$.
}
\end{figure}

A remarkable feature is that the hull function in both the $x$ and $y$ directions
becomes distorted by decreasing the anisotropy $\lambda$.
While the hull functions are smooth with larger anisotropies \cite{Yoshino10}, 
steplike structures appear when approaching the symmetric point $\lambda=1$.
The hull function at $\lambda=1$ appears to be nonanalytic: 
It is discontinuous at several points.
Figures~\ref{fig:dphi}(a) and \ref{fig:dphi}(b) show the discrete derivative of the hull functions 
$d \phit^\alpha(z)/dz \equiv [\phit^\alpha(z+2\pi/L) - \phit^\alpha(z)] / (2\pi/L)$.
It can be seen that sharp peaks emerge when approaching $\lambda=1$.
This is reminiscent of the critical behavior 
of the Aubry transition in the Frenkel-Kontorova model 
\cite{Peyrard83, Floria96, Yoshino10p}. 
Figure~\ref{fig:dphi}(b) shows the size dependence 
of the derivative at $\lambda=1$. 
As the size $L$ increases, not only do the peaks become sharper, 
but also the number of peaks increases reflecting higher harmonics.
We speculate that the hull function becomes discontinuous everywhere 
in the thermodynamic limit.

\subsection{Scaling behavior and diverging characteristic length}

\begin{figure}[ttt]
\begin{center}
\includegraphics[trim=0 40 120 20,scale=0.340,clip]{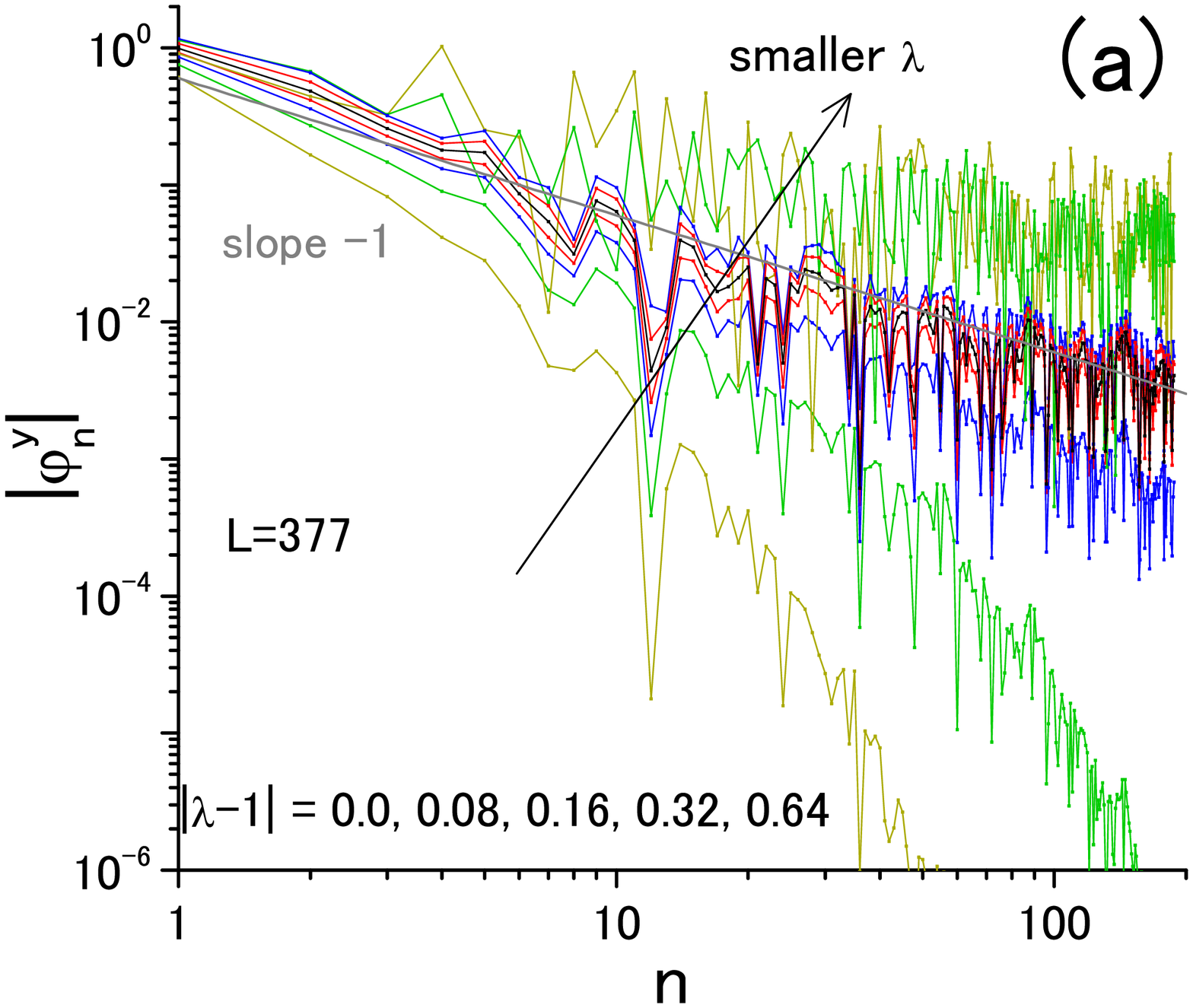}
\\
\includegraphics[trim=0 40 120 20,scale=0.340,clip]{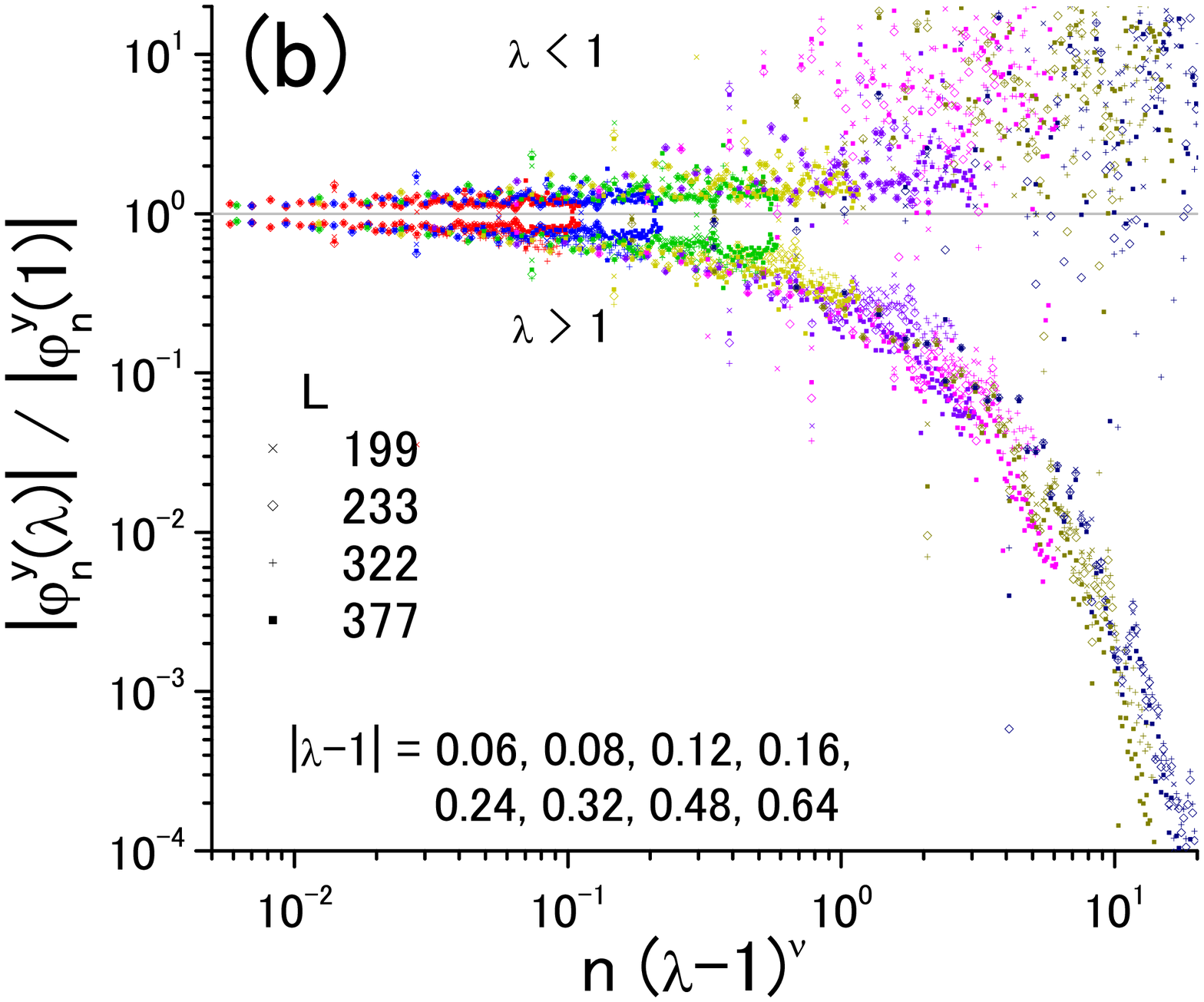}
\end{center}
\vspace{-5mm}
\caption{\label{fig:spectrum}
(Color online) 
Spectrum of the harmonics of the hull functions 
(a) before and (b) after scaling. We set $\nu=2.4$.
In the scaling, we do not use the data for $n>L/4$, 
which show a strong finite-size effect.
}
\end{figure}

Let us examine further the singular behavior around the symmetric point $\lambda=1.$
Figure~\ref{fig:spectrum}(a) shows 
the amplitude of the $n$th harmonic $n \qq$
[note that $\qq$ is set equal to $\qq_\mrm{GS}(\lambda=1)$ independently of $\lambda$]. 
At $\lambda=1$, $|\vphi^y_n|$ seems to decay as a 
power function of $n$ with exponent close to $-1$ 
although there are many dips and peaks.
In contrast, higher harmonics exponentially decrease 
above a certain characteristic scale $n^*(\lambda)$ for $\lambda>1$.
This cutoff indicates the smoothing of the hull functions. 
We found a scaling behavior 
\begin{equation}
\frac{|\vphi^{x,y}_n(\lambda)|}
{|\vphi^{x,y}_n(1)|}
= F^{x,y}_{\pm} \left( \frac{n}{n^*(\lambda)} \right)
\ \mrm{with}\ 
n^*(\lambda) \propto |\lambda-1|^{-\nu}
\label{eq:scaling}
\end{equation}
not only for $\lambda>1$, but also for $\lambda<1$, 
as shown in Fig.~\ref{fig:spectrum}(b).
The exponent $\nu$ is roughly estimated as $\nu \simeq 2.4$.
We confirmed that the same scaling works well for the FK model 
with $\nu=0.9874$ \cite{Coppersmith83, MacKay91}. 
This scaling means that a diverging number of higher harmonics 
come into play when approaching the symmetric point $\lambda=1$. 
We note that this observed singularity is present not only for $\qGS(1)$, 
but also for other $\qq$'s around $\qGS(1)$, 
i.e., for the continuous band of metastable states.

Initially, it would appear as if the characteristic length in Eq.~(\ref{eq:scaling}) 
were $1/n^*|\qq|$ and it went to zero, which is contrary to ordinary critical behavior. 
However, we speculate, inversely, that there is a diverging length scale for the following reasons.
On lattice systems, the component of the wave vector, $n q$, 
should be treated in the first Brillouin zone $(-1/2, 1/2)$ 
as $q_n^\mrm{FBZ} \equiv n q - \lfloor n q + 1/2 \rfloor$, 
where $\lfloor \cdots \rfloor$ is the floor function. 
For irrational $q$, $\{ q_n^\mrm{FBZ} \}$ behaves 
like a series of random numbers made by the linear congruential generator 
in a long span of $n$. 
Thus the maximum of $1/q_n^\mrm{FBZ}$ for $n \le n^*$ is roughly proportional to $n^*$.
If $q^x$ and $q^y$ are independent irrational numbers, 
the maximum of the vector length is proportional to $\sqrt{n^*}$. 
In addition, the number of harmonics in the finite-size system is bounded by $L$ 
when $q$ is approximated by an irreducible fraction $p/L$ 
(note that $q_{n+L}^\mrm{FBZ}=q_n^\mrm{FBZ}$). 
This also indicates that $n$ is a measure of the length.

The breakdown of the analyticity of the hull function means
that the sliding soft mode disappears; 
jamming occurs even in the direction where the hull function is gapless.
Indeed, of the transport properties of the present system it has been
found that the current-voltage characteristics exhibit a scaling feature 
around the symmetric point $\lambda=1$ \cite{Yoshino09}. 
Thus Eq.~(\ref{eq:scaling}) is regarded as a static signature 
of the jamming transition.

\section{Conclusion}

To summarize, we studied GSs and low lying metastable states of the IFJJA with anisotropic Josephson couplings.
We found the GS changes continuously with the variation of the anisotropy $\lambda$
of the Josephson coupling between the horizontal vortex stripes in the strong anisotropy limit and
nearly (but not exactly) diagonal vortex stripes at the isotropic point $\lambda=1$.

It is interesting to note again that the present system admits a continuous band of metastable
states around the GS, which is presumably the source of the glassiness in the present system. 
The analyticity breaking of the hull function approaching the isotropic point 
$\lambda \to 1$ implies jamming of the sliding vortices. 
In addition, the scaling behavior of the harmonic spectrum suggests 
the existence of diverging static length scale. 
It would be very interesting to study the properties of the
present system at finite temperatures from static and dynamic view-points.

\acknowledgments
This work was supported by 
a Grant-in-Aid for Scientific Research (C) (Grant No.~21540386),
a Grant-in-Aid for Scientific Research on Priority Areas 
``Novel States of Matter Induced by Frustration''(Grant No.~1905200*), 
and King Abdullah University of Science and Technology 
Global Research Partnership (Grant No.~KUK-I1-005-04).


%

\end{document}